\documentclass[english,aps,tightenlines,showpacs,showkeys,notitlepage]{revtex4-2}
\usepackage[T1]{fontenc}
\usepackage[latin9]{inputenc}
\usepackage{color}
\usepackage{babel}
\usepackage{amsmath}
\usepackage{amssymb}
\usepackage{graphicx}
\usepackage{setspace}
\usepackage{esint}
\usepackage[unicode=true,pdfusetitle,
 bookmarks=true,bookmarksnumbered=false,bookmarksopen=false,
 breaklinks=false,pdfborder={0 0 1},backref=false,colorlinks=true]
 {hyperref}
\hypersetup{
 pdfborderstyle=}

\makeatletter
\usepackage{babel}

\makeatother

\begin{document}
\title{Exact vacuum FLRW solutions in $q$-deformed Brans-Dicke cosmology}
\author{Salih Kibaro\u{g}lu$^{1}$}
\email{salihkibaroglu@maltepe.edu.tr}

\author{Mustafa Senay$^{2}$}
\email{msenay@bartin.edu.tr}

\date{\today}
\begin{abstract}
We study a $q$-deformed extension of Brans-Dicke gravity in a spatially
flat Friedmann-Lemaître-Robertson-Walker space-time. The deformation
enters through a coupling function that modifies the effective gravitational
strength and leads to generalized Friedmann equations. In the matter-free
sector, we obtain exact analytic solutions for the scale factor and
the Brans-Dicke scalar field, and recast the scalar contribution as
an effective fluid. We show that the corresponding equation-of-state
parameter and the deceleration parameter are constants and depend
only on the Brans-Dicke coupling $\omega$ and the deformation function,
allowing the scalar sector to mimic radiation-, matter-, or dark-energy-like
behavior for a restricted region of parameter space.
\end{abstract}
\affiliation{$^{1)}$Maltepe University, Faculty of Engineering and Natural Sciences,
34857, Istanbul, Türkiye}
\affiliation{$^{2)}$Department of Medical Services and Techniques, Vocational
School of Health Services, Bart\i n University, 74100, Bart\i n, Türkiye}
\maketitle

\section{Introduction}

The gravitational interaction may receive corrections beyond General
Relativity when new degrees of freedom, thermodynamic arguments, or
quantum/statistical effects are taken seriously. A broad class of
such modifications is provided by scalar-tensor theories \cite{Nojiri:2011UnifiedCosmic,Capozziello:2011Extended,Clifton:2011Modified,Nojiri:2017Modified},
where the effective gravitational coupling becomes dynamical. Brans-Dicke
(BD) theory \cite{Brans:1961sx} is the prototypical example, and
has been applied to cosmology and compact objects in many contexts,
including black holes \cite{Hawking:1972BlackHolesBD,Sotiriou:2011Blackholes,Adak:2018Non-Riemannian},
interactions with matter fields \cite{Dereli:1982WEYL}, time-varying
gravitational coupling and dark sector phenomenology \cite{Damour:1990DarkMatter,Setare:2006TheHolographicDE,Shaposhnikov:2009Scale,Karananas:2016Scale},
scale-invariant inflationary scenarios \cite{Ferreira:2018Inflation,Casas:2019Scale,Ghilencea:2021GaugingScale},
non-Riemannian formulations \cite{Adak:2018Non-Riemannian,Dereli:2019GravitationalPlane},
gravity waves \cite{Dereli:2019GravitationalPlane}, and parameter-space
studies of $\omega$ and observational constraints \cite{Dereli:2022DarkRange}.

In parallel, the thermodynamic interpretation of gravitation triggered
by black hole entropy and Hawking radiation \cite{Bekenstein1973,Hawking:1975vcx}
and later developed in various forms \cite{Jacobson:1995Thermodynamics,Padmanabhan:2005Gravity,Padmanabhan:2010Thermodynamical}
has motivated derivations of cosmological equations from horizon thermodynamics
\cite{Cai:2005FirstLaw,Akbar:2007Thermodynamic,Cai:2007Unified,Nojiri:2022Modified,Nojiri:2022Early,Nojiri:2024Different}
and, more broadly, emergent/entropic gravity ideas \cite{Verlinde:2010hp,Verlinde:2016toy}.
Within this viewpoint, it is natural to ask how nonstandard microscopic
statistics feeds into macroscopic gravitational dynamics.

Deformed-statistics frameworks (including $q$-deformations) provide
a systematic way to encode nonstandard microscopic behavior and have
been used to obtain modified Einstein/Friedmann equations and related
cosmological consequences \cite{Senay:2018xaj,Kibaroglu:2018mnx,Kibaroglu:2019odt,Senay_2021,SENAY2021136536,Senay_2024,Senay:2024Implications}.
However, the interplay between (i) a genuinely scalar-tensor gravitational
sector in the BD sense and (ii) $q$-deformation emerging from an
entropic-gravity construction has not been fully explored at the level
of explicit cosmological solutions and their effective fluid interpretation.
In particular, it is not obvious a priori whether the resulting scalar
sector can mimic standard cosmological phases (radiation/matter/dark
energy), or whether it introduces qualitatively new expansion histories. 

In this work, we study a $q$-deformed extension of BD gravity - obtained
previously from Verlinde-type entropic arguments \cite{Kibaroglu:2025QBD}
- in a spatially flat Friedmann-Lemaître-Robertson-Walker (FLRW) background.
The deformation enters through a coupling function $\alpha\left(z,q\right)$
and modifies the effective gravitational coupling. The paper is organized
as follows. In Sec. \ref{sec:Scalar-tensor-theory-1} we review the
$q$-deformed BD field equations obtained from the entropic-gravity
setup. In Sec. \ref{sec:Cosmological-analysis} we derive the corresponding
FLRW equations and study their exact solutions in the matter-free
sector, together with the effective EoS and deceleration parameter.
We conclude in Sec.\ref{sec:Conclusion}.

Notations and conventions: In this work, we adopt the mostly negative
metric signature, such that the four-dimensional Minkowski metric
takes the form $\text{diag}(1,-1,-1,-1)$. Throughout the paper, we
work in natural units by setting the fundamental physical constants
to unity: the speed of light $c=1$, and Boltzmann's constant $k_{B}=1$.
The four-dimensional space-time indices are represented by Roman letters
such as $a,b,c,...=0,1,2,3$.

\section{\label{sec:Scalar-tensor-theory-1}$q$-Deformed Brans-Dicke theory }

Here we briefly summarize the formulation introduced in Ref.\cite{Kibaroglu:2025QBD},
where BD gravity is extended by $q$-deformed statistics within Verlinde's
entropic-gravity framework.

\subsection{$q$-deformed background}

The deformation is incorporated in the entropy-area relation and therefore
in the local Unruh temperature on the holographic screen. Using the
conformal transformation $\tilde{g}_{ab}=\psi g_{ab}$, with $\psi$
the BD scalar, the local temperature measured by an accelerated observer
becomes \cite{Algin:2012df,Algin:2016cuo,Algin:2016df}

\begin{equation}
T=\alpha\left(z,q\right)\frac{\hbar}{2\pi}e^{\phi}N^{a}\left(a_{a}+\frac{1}{2}\nabla_{a}\log\psi\right),\label{eq:q-temperature1}
\end{equation}
where $\phi=\frac{1}{2}\log\sqrt{-\xi^{a}\xi_{a}}$ is the relativistic
gravitational potential ($\xi^{a}$ is a time-like Killing field),
accordingly, the exponential factor $e^{\phi}$ represents the redshift
between local and asymptotic observers, with $\phi=0$ taken as the
reference point at spatial infinity, $a^{b}=-\nabla^{b}\phi$ is the
Jordan-frame acceleration, and $N^{a}$ is the outward unit normal.
The function $\alpha\left(z,q\right)$ encodes the statistical deformation:

\begin{equation}
\alpha\left(z,q\right)=\frac{5N}{2}\left[\frac{5}{2}\frac{f_{5/2}(z,q)}{f_{3/2}(z,q)}-\ln z\right],\label{eq:alpha}
\end{equation}
where $0<q<1$ measures the deformation strength, $z=\exp\left(\mu/T\right)$
is the fugacity ($\mu$ is the chemical potential) and $N$ is the
total particle number,
\begin{equation}
N=\sum_{i}\frac{1}{|\ln q|}\left|\ln\left(\frac{|1-ze^{-x_{i}}|}{1+qze^{-x_{i}}}\right)\right|,\label{eq:totalnumber1}
\end{equation}
where $x_{i}=\beta\epsilon_{i}$ is a dimensionless variable representing
the energy scale of the system, introduced to simplify the evaluation
of thermostatistical functions and integrals. Here $\beta=1/k_{B}T$
denotes the inverse thermal energy, and $\epsilon_{i}$ represents
the kinetic energy of a particle in the $i$-th energy state. The
function $f_{n}(z,q)$ denotes the generalized Fermi--Dirac function,
\begin{eqnarray}
f_{n}(z,q) & = & \frac{1}{\Gamma(n)}\int_{0}^{\infty}\frac{x^{n-1}}{|\ln q|}\left|\ln\left(\frac{|1-ze^{-x}|}{1+qze^{-x}}\right)\right|\text{d}x\nonumber \\
 & = & \frac{1}{|\ln q|}\left[\sum_{l=1}^{\infty}(-1)^{l-1}\frac{(zq)^{l}}{l^{n+1}}-\sum_{l=1}^{\infty}\frac{z^{l}}{l^{n+1}}\right],\label{eq:q-FD}
\end{eqnarray}
where $\Gamma\left(n\right)=\int_{0}^{\infty}x^{n-1}e^{-x}dx$ is
the Gamma function for $n>0$ and $x=\beta\epsilon$. 

For Eqs.\eqref{eq:totalnumber1} and \eqref{eq:q-FD}, different regimes
of $x$ characterize the relative contribution of energy states. In
the limit $x\ll1$, low-energy states dominate and the deformation
effects associated with $q$ become more pronounced. In contrast,
for $x\gg1$ the factor $e^{-x}$ strongly suppresses high-energy
contributions. The intermediate regime $x\approx1$ corresponds to
the characteristic thermal scale where particle energies are comparable
to the thermal energy and therefore contribute most significantly
to the generalized Fermi--Dirac integrals appearing in $\alpha\left(z,q\right)$.
In this work, we focus on this regime in order to capture the dominant
thermodynamic behavior of the system rather than the extreme limits.

\begin{figure}
\includegraphics[width=9cm]{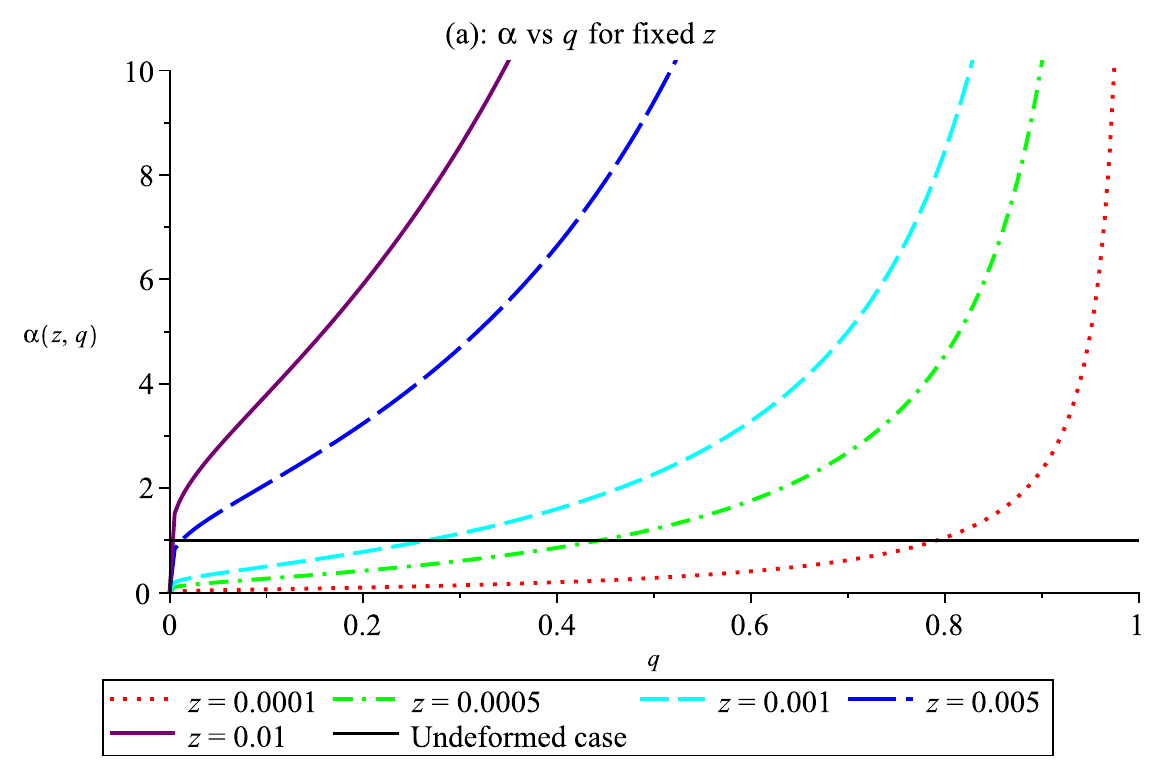}\includegraphics[width=9cm]{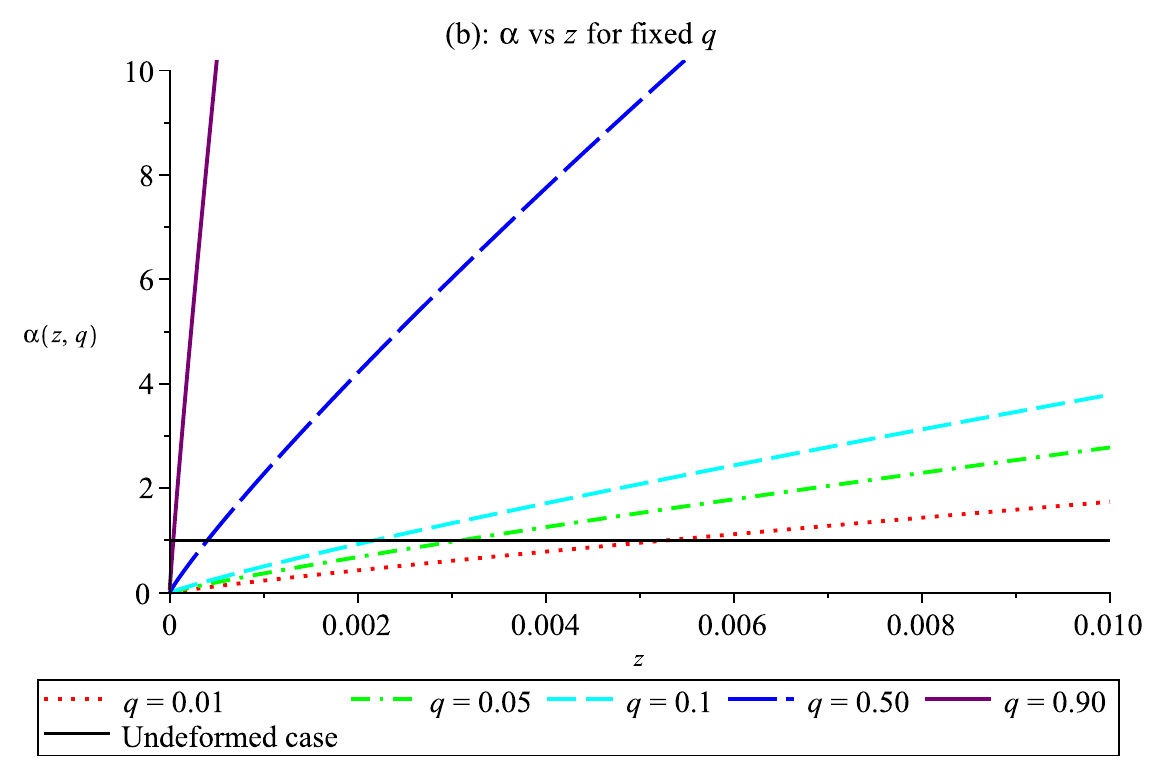}

\caption{\label{fig: alpha}(color online) Panel (a) shows $\alpha(z,q)$ as
a function of the deformation parameter $q$ for fixed values of the
fugacity $z$. Panel (b) shows $\alpha(z,q)$ as a function of the
fugacity $z$ for fixed values of the deformation parameter $q$.
The undeformed case $\alpha(z,q)=1$ is shown as the black horizontal
reference line.}
\end{figure}

In Fig. \ref{fig: alpha}, we illustrate the behavioral analysis of
the statistical deformation function $\alpha\left(z,q\right)$ with
respect to various values of the deformation parameter $q$ and fugacity
$z$ in the regime $x\approx1$. Furthermore, $\alpha\left(z,q\right)$
spans a wide range, from values close to zero to very large magnitudes
($\sim10^{7}$) as both $q$ and $z$ approach unity. This wide variability
allows $\alpha\left(z,q\right)$ to be interpreted as an effective
cosmological free parameter in our model.

\subsection{Generalized gravitational field equations}

By invoking the equipartition theorem on the holographic screen, we
derived the total mass $M$ which can be seen a modified version of
Gauss's law under scalar-tensor gravity, 
\begin{equation}
M=\frac{\alpha\left(z,q\right)}{4\pi}\oint_{\partial\Sigma}e^{\phi}N^{a}\left(a_{a}+\frac{1}{2}\nabla_{a}\log\psi\right)\psi\text{d}A,\label{eq:q-mass}
\end{equation}
where $\mathrm{d}A$ is the surface area element on $\partial\Sigma$.
Using Stokes' theorem and matching with the Komar definition gives
the generalized field equations

\begin{eqnarray}
R_{ab}-\frac{1}{2}g_{ab}R & = & \frac{8\pi}{\alpha\left(z,q\right)\psi}\left(T_{ab}^{M}+T_{ab}^{\psi}\right),\label{eq: q-Einstein_BD}
\end{eqnarray}
with matter contribution $T_{ab}^{M}$ and scalar contribution

\begin{equation}
T_{ab}^{\psi}=\frac{1}{8\pi}\left[\frac{\omega}{\psi}\left(\partial_{a}\psi\partial_{b}\psi-\frac{1}{2}g_{ab}\partial_{c}\psi\partial^{c}\psi\right)+\alpha\left(z,q\right)\left(\nabla_{a}\nabla_{b}-g_{ab}\square\right)\psi\right],
\end{equation}
where $\square\equiv g^{ab}\nabla_{a}\nabla_{b}$ denotes the d'Alembert
operator and $\omega$ is a dimensionless coupling parameter analogous
to the BD parameter. Additionally, Eq. \eqref{eq: q-Einstein_BD}
can be written as $G_{ab}=8\pi G_{{\rm eff}}\left(T_{ab}^{M}+T_{ab}^{\psi}\right)$
with 
\begin{equation}
G_{{\rm eff}}=\left[\alpha\left(z,q\right)\psi\right]^{-1},
\end{equation}
which makes explicit how the deformation and scalar sector jointly
renormalize the gravitational strength. However, in our equations,
we prefer to use the exact form to see the effect of the deformation
parameter explicitly.

Consistency limits are immediate: (i) for $\alpha\left(z,q\right)=1$
one recovers standard BD theory \cite{Brans:1961sx}; (ii) for $\psi=1$
one obtains the $q$-deformed Einstein equations \cite{Senay:2018xaj,Kibaroglu:2018mnx,Kibaroglu:2019odt,SENAY2021136536};
and (iii) imposing both $\alpha=1$ and $\psi=1$ yields Einstein
gravity.

In Sec. \ref{sec:Cosmological-analysis}, we specialize these equations
to a spatially flat FLRW background.

\section{\label{sec:Cosmological-analysis}Cosmological analysis}

In general, for most cosmological models, the metric is assumed to
be a flat FLRW metric which is based on Einstein's equations of general
relativity. This framework facilitates the description of the universe's
evolution under the assumptions of homogeneity and isotropy on large
scales. Accordingly, we begin with the following line element:

\begin{eqnarray}
ds^{2} & = & dt^{2}-a\left(t\right)^{2}\left(dx^{2}+dy^{2}+dz^{2}\right),\label{eq: FLRW metric}
\end{eqnarray}
where $a\left(t\right)$ is the scale factor. Consistent with our
assumptions, we consider the scalar field as a function of cosmic
time, denoted as $\psi\left(x\right)\rightarrow\psi\left(t\right)$.
Accordingly, the energy-momentum tensors $T_{ab}^{M}$ and $T_{ab}^{\psi}$
can be consistently interpreted as those of classical perfect fluids,
as they take the standard diagonal form in a comoving frame:

\begin{equation}
T_{\,\,\,\,\,\,\,b}^{M\,a}=diag\left(\rho_{M},-P_{M},-P_{M},-P_{M}\right),
\end{equation}
and

\begin{equation}
T_{\,\,\,\,\,\,\,b}^{\psi\,a}=diag\left(\rho_{\psi},-P_{\psi},-P_{\psi},-P_{\psi}\right).
\end{equation}
where $\rho_{M}$, $\rho_{\psi}$ denote the energy densities and
$P_{M}$, $P_{\psi}$ represent the corresponding isotropic pressures
of the matter and scalar field components, respectively. These tensors
are consistent with the general form of a perfect fluid energy-momentum
tensor, $T_{ab}=\left(\rho+P\right)u_{a}u_{b}-Pg_{ab}$ where $u_{a}$
is the fluid's four-velocity. Furthermore, each fluid component can
be characterized by an EoS parameter defined as $W_{i}=P_{i}/\rho_{i}$,
with $i=M,\psi$, which encapsulates the thermodynamic relation between
pressure and energy density. This framework allows for a straightforward
cosmological interpretation of both matter and scalar field contributions
in terms of perfect fluids.

Having established the necessary background, we can now proceed to
derive the Friedmann equations. The following equation can be obtained
by focusing on the $\left(0,0\right)$ component of Eq.(\ref{eq: q-Einstein_BD}),

\begin{equation}
3H^{2}=\frac{8\pi}{\alpha\left(z,q\right)\psi}\rho_{M}+\frac{\omega}{2}\frac{\dot{\psi}^{2}}{\alpha\left(z,q\right)\psi^{2}}-3H\frac{\dot{\psi}}{\psi},\label{eq: F1}
\end{equation}
where $H\left(t\right)=\dot{a}\left(t\right)/a\left(t\right)$ is
the Hubble parameter. The $\left(i,i\right)$ components of Eq.(\ref{eq: q-Einstein_BD})
lead to,

\begin{equation}
2\dot{H}+3H^{2}=-\frac{8\pi}{\alpha\left(z,q\right)\psi}P_{M}-\frac{\omega}{2\alpha\left(z,q\right)}\frac{\dot{\psi}^{2}}{\psi^{2}}-2H\frac{\dot{\psi}}{\psi}-\frac{\ddot{\psi}}{\psi},\label{eq: F2}
\end{equation}
where the dot denotes the derivative with respect to the cosmic time
($d/dt$) and the pressure $P_{\psi}$ and energy density $\rho_{\psi}$
parameters can be written as,

\begin{equation}
\rho_{\psi}=\frac{1}{8\pi}\left[\frac{\omega}{2}\frac{\dot{\psi}^{2}}{\psi}-3\alpha\left(z,q\right)H\dot{\psi}\right],\label{eq: rho_psi}
\end{equation}

and

\begin{equation}
P_{\psi}=\frac{1}{8\pi}\left[\frac{\omega}{2}\frac{\dot{\psi}^{2}}{\psi}+\alpha\left(z,q\right)\left(2H\dot{\psi}+\ddot{\psi}\right)\right].\label{eq: P_psi}
\end{equation}
Then, by making use of Eqs.(\ref{eq: F1}) and (\ref{eq: F2}), the
acceleration equation can be obtained as,

\begin{equation}
\frac{\ddot{a}}{a}=-\frac{4\pi}{3\alpha\left(z,q\right)\psi}\left[3\left(P_{M}+P_{\psi}\right)+\rho_{M}+\rho_{\psi}\right].\label{eq: Friedmann_acceleration}
\end{equation}
These are the deformed Friedmann equations in the context of $q$-deformed
BD gravitation theory. For completeness, if we combine Eqs.(\ref{eq: F1})
and (\ref{eq: Friedmann_acceleration}), we also derive the time derivative
of the Hubble parameter as 
\begin{equation}
\dot{H}=-\frac{4\pi}{\alpha\left(z,q\right)\psi}\left(P_{M}+\rho_{M}\right)-\frac{\omega}{2\alpha\left(z,q\right)}\left(\frac{\dot{\psi}}{\psi}\right)^{2}+\frac{H}{2}\frac{\dot{\psi}}{\psi}-\frac{\ddot{\psi}}{2\psi}.
\end{equation}
This equation provides a detailed description of the evolution of
the Hubble parameter, including contributions from matter, the scalar
field, and the deformation function $\alpha\left(z,q\right)$. 

\subsection{\label{subsec:Solutions-for-the}Solutions for the Friedmann equations}

We now turn to the analysis of the Friedmann equations in the absence
of ordinary matter. In this vacuum configuration, the cosmic dynamics
are governed entirely by the modified gravitational sector, namely
the scalar field and the $q$-deformation effects, without contributions
from standard matter or radiation components. This setting enables
us to isolate and clearly identify the pure impact of the $q$-deformed
scalar--tensor structure on the cosmological evolution. Under these
assumptions, solving Eq.\eqref{eq: F1} for the scalar field $\psi\left(t\right)$
yields the following relation:

\begin{equation}
\psi\left(t\right)^{\pm}=\psi_{0}a\left(t\right)^{\frac{\mathcal{A}^{\pm}+3\alpha}{\omega}},\label{eq: sol_psi}
\end{equation}
where $\psi_{0}$ is a constant and 
\begin{eqnarray}
\mathcal{A}^{\pm} & = & \pm\sqrt{3\alpha\left(2\omega+3\alpha\right)},
\end{eqnarray}
is introduced for simplicity (here $\alpha$ is the shorter representation
of the deformation function $\alpha\left(z,q\right)$). Here, it is
important to note that in order to avoid complex values, the parameter
$\omega$ have to satisfy the following condition 
\begin{eqnarray}
\omega & \geq & -\frac{3}{2}\alpha.\label{eq: cond_omega_alpha}
\end{eqnarray}
In analogy with BD theory, we have demonstrated that the scalar field
evolves as a power-law function of the scale factor $a\left(t\right)$,
consistent with theoretical expectations (see \cite{Pimentel:1985exact,Banerjee:2007HolographicDE_BD,Sheykhi:2010Interacting,Khatri:2023ojm}).
Notably, in \cite{Peracaula:2018BD_Higgs}, this power-law behavior
of the BD scalar was assumed to explore the dynamics of BD cosmology
with a cosmological term. Furthermore, it has been observed that adopting
this power-law form for the BD scalar significantly enhances the fit
to cosmological data \cite{deCruzPerez:2018BD_cosm_mimic}.

Moreover, substituting Eq.(\ref{eq: sol_psi}) into Eq.(\ref{eq: F2})
and solving with respect to $a\left(t\right)$ we get 
\begin{equation}
a\left(t\right)^{\pm}=\left[\mathcal{B}^{\pm}\left(\psi_{0}t+C_{1}\right)\right]^{\frac{1}{\mathcal{B}^{\pm}}},\label{eq: sol_a}
\end{equation}
where $C_{1}$ is an integral constant and the expression $\mathcal{B}^{\pm}$
is introduced for simplicity as 
\begin{eqnarray}
\mathcal{B}^{\pm} & = & \frac{5\left(\omega+\frac{6}{5}\alpha\right)\mathcal{A}^{\pm}-6\left(\omega+2\alpha\right)\left(\omega+\frac{3}{2}\alpha\right)}{\omega\left(\mathcal{A}^{\pm}-3\alpha-2\omega\right)}.
\end{eqnarray}
Subsequently, using Eq.(\ref{eq: sol_a}) in Eq.(\ref{eq: sol_psi})
we find; 
\begin{equation}
\psi^{\pm}\left(t\right)=\psi_{0}\left[\mathcal{B}^{\pm}\left(\psi_{0}t+C_{1}\right)\right]^{\frac{-\mathcal{A}^{\pm}+3\alpha}{\omega\mathcal{B}^{\pm}}}.\label{eq: sol_psi_exact}
\end{equation}

As a result, we have derived the exact expression for the scalar field
as a function of the coupling constant $\omega$ and the deformation
function $\alpha$, along with certain integration constants.

In the subsequent subsection, we extend the analysis to encompass
various types of cosmic evolution, characterized by the BD parameter
$\omega$, including scenarios such as radiation-dominated or dark
energy-like evolution.

\subsection{\label{subsec:Analyzing-the-equation}Analyzing the equation of state
function}

In this part, we investigate the equation of state (EoS) function
within the context of the proposed q-deformed scalar--tensor framework.
This analysis is crucial for understanding the behavior of cosmological
fluids under the modified gravitational dynamics introduced by $q$-deformation.
In the absence of ordinary matter, the scalar field $\psi$ effectively
behaves as a kinetic fluid component. Importantly, in the vacuum configuration
considered here, the scalar sector satisfies an independent conservation
equation derived from the modified field equations, namely

\begin{equation}
\dot{\rho}_{\psi}+3H\left(1+W_{\psi}^{\pm}\right)\rho_{\psi}=0,
\end{equation}
where the EoS parameter is defined as $W_{\psi}^{\pm}=\frac{P_{\psi}}{\rho_{\psi}}$
. Since the modified Friedmann equations can be consistently written
in an effective GR-like form with a self-conserved scalar fluid, the
standard relation 
\begin{equation}
W_{\psi}^{\pm}=-1-\frac{2\dot{H}}{3H^{2}},
\end{equation}
remains valid within the present framework. By applying this expression
together with Eq.\eqref{eq: sol_a}, we obtain

\begin{equation}
W_{\psi}^{\pm}=-1+\frac{2}{3}\mathcal{B}^{\pm}.\label{eq: EoS}
\end{equation}
We observe that, in this specific case, the EoS parameter is time-independent
and depends solely on the deformation function $\alpha$ and the BD
parameter $\omega$. This result indicates that cosmic evolution is
governed predominantly by the intrinsic parameters of the modified
gravitational theory rather than by dynamical matter-energy interactions.
Such behavior can lead to scenarios where acceleration or deceleration
phases are determined inherently by the gravitational model rather
than by evolving matter components.

\begin{figure}
\includegraphics[width=15cm]{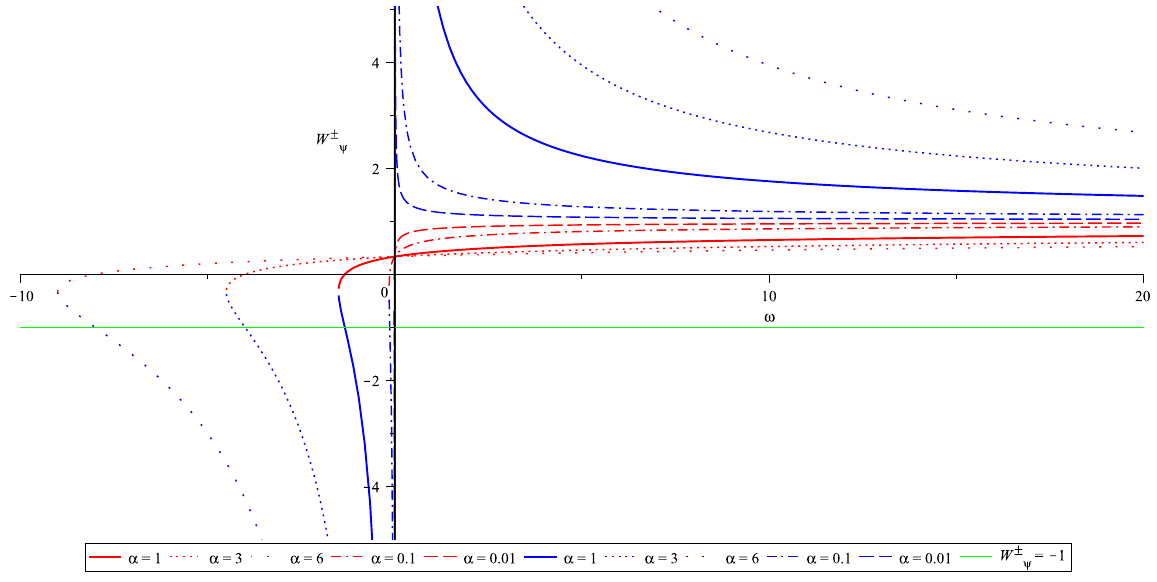}

\caption{\label{fig: EoS}(color online) The equations of state functions $W_{\psi}^{\pm}$
for the scalar field $\psi$ versus the parameter $\omega$ for various
values of the deformation function $\alpha$. The red and blue lines
represent $W_{\psi}^{+}$ and $W_{\psi}^{-}$ functions, respectively.
The green line corresponds to the dark energy dominated era ($W_{\psi}^{\pm}=-1$).}
\end{figure}

Figure \ref{fig: EoS} illustrates the behavior of the EoS parameter
$W_{\psi}^{\pm}$ given in Eq.\eqref{eq: EoS} as a function of the
BD coupling parameter $\omega$ for several representative values
of the deformation function $\alpha$, where $\alpha=1$ corresponds
to the undeformed (standard BD) limit. This comparison allows us to
explicitly identify the deviations induced by $q$-deformation relative
to the conventional BD cosmology (see, e.g., \cite{Lee:2011EoS_BD}).
The analysis reveals that the following conditions are satisfied:
\begin{enumerate}
\item In the limit of $\omega\rightarrow0$; $W_{\psi}^{-}$ becomes an
undefined function while $W_{\psi}^{+}$ stabilizes at $\frac{1}{3}$, 
\item For the positive values of $\omega$; $1\leq W_{\psi}^{-}\leq\infty$
and $\frac{1}{3}<W_{\psi}^{+}\leq1$, 
\item For the negative values of $\omega$; $-\infty\leq W_{\psi}^{-}\leq-\frac{1}{3}$
and $-\frac{1}{3}\leq W_{\psi}^{+}<\frac{1}{3}$. 
\end{enumerate}

Physically, the plot indicates that the $q$-deformation (encoded
in $\alpha$) acts as a control parameter that shifts the scalar sector
between standard cosmological behaviors: $W_{\psi}\simeq1/3$ (radiation-like),
$W_{\psi}\simeq0$ (matter-like), and $W_{\psi}\simeq-1$ (dark-energy-like).
In particular, the $W_{\psi}^{-}$ branch accesses the negative-pressure
domain ($W_{\psi}<-1/3$) required for accelerated expansion, whereas
the $W_{\psi}^{+}$ branch remains non-accelerating for the same parameter
ranges.

\begin{figure}
\includegraphics[width=9cm]{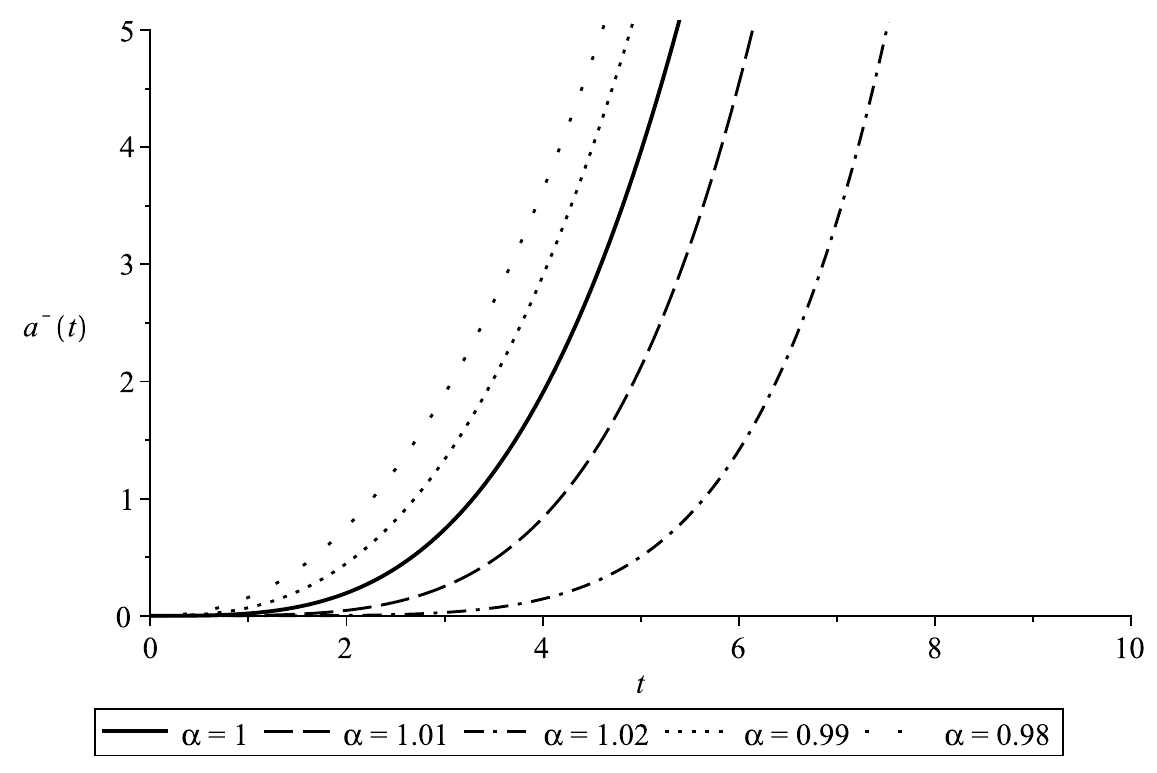}\includegraphics[width=9cm]{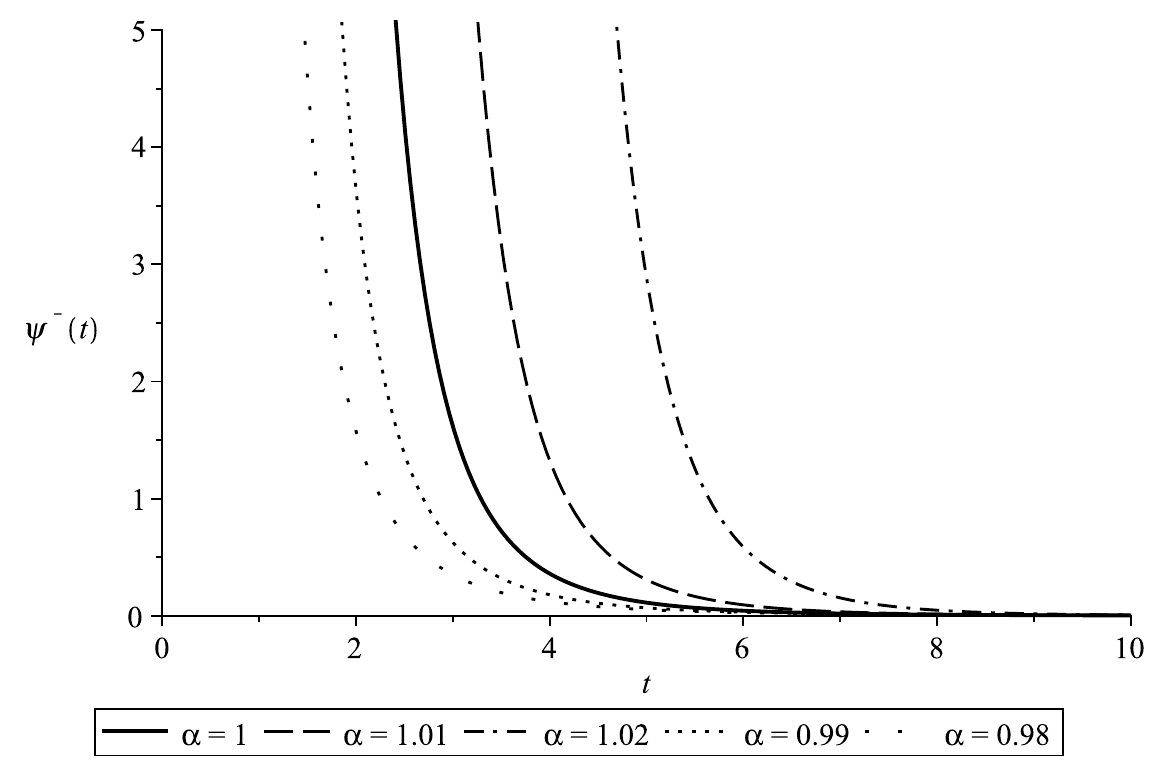}

\caption{\label{fig: a(t)_and_psi(t)} The time evolution of the scale factor
$a\left(t\right)^{-}$ and the scalar field $\psi\left(t\right)^{-}$
versus time for various values of the deformation function $\alpha$.
Here the constants are set to be $\psi_{0}=1$, $C_{1}=0$, $\omega=-1.4$
and $\alpha=1$ (solid line) which corresponds to the non-deformed
case.}
\end{figure}

Accordingly, the plots in Figure \ref{fig: a(t)_and_psi(t)} exemplify
the accelerated dynamics of the universe and illustrate the impact
of the deformation function $\alpha$ on both the scale factor $a\left(t\right)^{-}$
and the scalar field $\psi\left(t\right)^{-}$ within the framework
of $q$-deformed scalar-tensor gravity under specific conditions.

\subsection{Phenomenology and viability in the matter-free sector}

The matter-free solution \eqref{eq: sol_a} implies a power-law expansion
$a\left(t\right)^{\pm}\propto\left(\psi_{0}t+C_{1}\right)^{1/\mathcal{B}^{\pm}}$,
which yields the Hubble parameter 
\begin{equation}
H\left(t\right)^{\pm}=\frac{\dot{a}}{a}=\frac{1}{\mathcal{B}^{\pm}\left(t+C_{1}/\psi_{0}\right)},
\end{equation}
then the deceleration parameter becomes

\begin{equation}
q_{\psi}^{\pm}=-\frac{\ddot{a}a}{\dot{a}^{2}}=\mathcal{B}^{\pm}-1.\label{eq: deceleration}
\end{equation}
where $q_{\psi}^{\pm}$ denotes the deceleration parameter and should
not be confused with the deformation parameter $q$. This equation
shows that $q_{\psi}^{\pm}$ is constant and independent of cosmic
time. Typically, in cosmological models, deceleration parameter evolves
over time, reflecting different expansion phases, such as an early
decelerating phase followed by late-time acceleration. However, the
solution without matter sources, $q_{\psi}^{\pm}$ is solely determined
by the deformation function $\alpha$ and the BD parameter $\omega$,
implying that the expansion behavior is fixed once these parameters
are set. 

Therefore, accelerated expansion requires $\mathcal{B}<1$ (equivalently
$W_{\psi}<-1/3$ via Eq.\eqref{eq: EoS}), which occurs only for the
negative-sign branch in Eq.\eqref{eq: sol_psi} for appropriate values
of $\left(\omega,\alpha\right)$. Similar to the pure BD gravity,
this solution requires also $\omega<0$ for the dark energy era. On
the other hand, the Solar System and cosmological observations constrain
the BD parameter to be large and positive (typically $\omega\gg1$)
\cite{Liddle:1998Radiation,Nagata:2004WMAPConstraints,Acquaviva:2005StructureFormation,Avilez:2014CosmologicalBD}.
In conventional BD theory, such large values generally suppress scalar-field--driven
acceleration in the absence of a potential or additional matter sources.
Therefore a full assessment of viable cosmological solution may include
matter, possible scalar self-interactions, additional scalar potential
$V\left(\psi\right)$ and field-dependent $\omega\rightarrow\omega\left(\psi\right)$
\cite{Alsing:2012Gravitational,Kofinas:2016ModifiedBD}.

A basic viability requirement is the positivity of the effective gravitational
coupling in Eq.(\ref{eq: q-Einstein_BD}), i.e. 
\begin{equation}
\alpha\left(z,q\right)\,\psi>0,
\end{equation}
together with the reality condition (\ref{eq: cond_omega_alpha}).
Moreover, the exact solution (\ref{eq: sol_psi_exact}) directly determines
the time variation of $G_{\mathrm{eff}}$ through $\dot{G}_{\mathrm{eff}}/G_{\mathrm{eff}}=-\dot{\psi}/\psi$
when $\alpha$ is effectively constant. This provides a simple phenomenological
handle on the model and motivates a future, more detailed confrontation
with local and cosmological constraints.

\begin{figure}
\includegraphics[width=15cm]{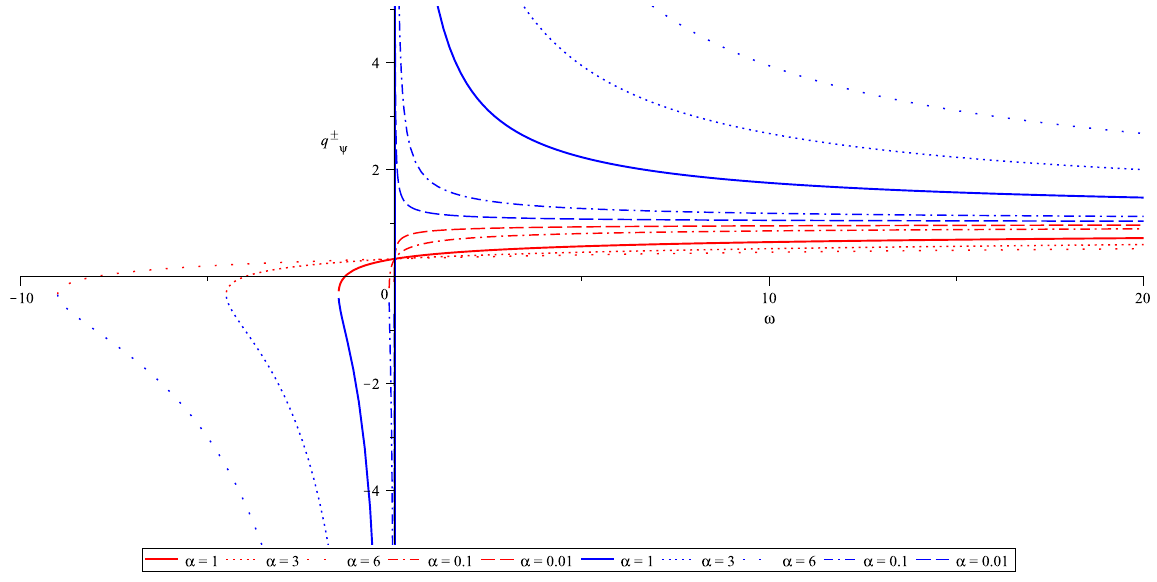}

\caption{\label{fig: deceleration}(color online) The deceleration parameter
$q_{\psi}^{\pm}$ versus the parameter $\omega$ for various values
of the deformation function $\alpha$. The red and blue lines represent
$q_{\psi}^{+}$ and $q_{\psi}^{-}$ functions, respectively, where
$\alpha=1$ (solid line) which corresponds to the non-deformed case,
$\alpha=3$ (dotted line), $\alpha=6$ (spaced dotted line), $\alpha=0.1$
(dash-dot line) and $\alpha=0.01$ (dashed line). }
\end{figure}

Fig. \ref{fig: deceleration} shows the deceleration parameter $q_{\psi}^{\pm}$
as a function of the BD coupling $\omega$ for several fixed values
of the deformation function $\alpha$ and therefore provides a compact
parameter-space diagnostic for the vacuum solutions: the two branches
behave qualitatively differently. The ($+$) branch typically remains
decelerating ($q_{\psi}^{+}>0$) and tends to an asymptotic value
as $\omega$ increases, while the ($-$) branch can enter an accelerating
regime ($q_{\psi}^{-}<0$) for an allowed range of $(\omega,\alpha)$.

\section{Conclusion\label{sec:Conclusion}}

We have investigated a $q$-deformed generalization of BD gravity,
motivated by entropic-gravity considerations, in a spatially flat
FLRW background. We derived the modified Friedmann equations, Eqs.\eqref{eq: F1}
and \eqref{eq: F2}, and obtained exact vacuum solutions for the scale
factor, Eq.\eqref{eq: sol_a}, and scalar field, Eq.\eqref{eq: sol_psi_exact}.
These solutions reduce to the standard BD expressions in the limit
$\alpha\left(z,q\right)\to1$ (see Fig. \ref{fig: alpha}).

Interpreting the scalar sector as an effective fluid, we found that
the EoS parameter $W_{\psi}^{\pm}$ is time independent and is controlled
only by the deformation function $\alpha\left(z,q\right)$ and the
BD coupling $\omega$. Depending on the branch and parameter choices,
the scalar sector can mimic radiation-, matter-, quintessence-, or
phantom-like behavior (see Fig. \ref{fig: EoS}).

For the vacuum solutions studied here, accelerated expansion occurs
only for the minus branch (see Eq.\eqref{eq: sol_psi}) in a restricted
region of parameter space which requires $\omega<0$ (see Figs. \ref{fig: EoS}
and \ref{fig: a(t)_and_psi(t)}) in accordance with the pure BD gravity.
We also found that the deceleration parameter $q_{\psi}^{\pm}$ is
constant, Eq.\eqref{eq: deceleration} (Fig. \ref{fig: deceleration}). 

However, standard BD cosmology without additional ingredients (e.g.,
a scalar potential or extra matter couplings) is known to have difficulty
generating late-time acceleration for observationally allowed large
positive values of $\omega$. Therefore, the matter-free solutions
presented here should be viewed primarily as an analytic toy model
that isolates the role of $\alpha\left(z,q\right)$. Constructing
a fully realistic cosmology consistent with current bounds ($\omega\gg1$)
likely requires additional sectors, which we leave for future work. 

The present framework also exhibits a close correspondence with effective
modified-gravity theories. In particular, since $f(R)$ gravity can
be reformulated as a scalar-tensor theory equivalent to a BD-type
model, the obtained $q$-deformed cosmological equations may likewise
be interpreted within an effective gravitational theory \cite{Nojiri:2022FromNonextensive,Nojiri:2025TheCorrespondence},
in which the generalized cosmological dynamics arise from modified
entropy functionals. However, unlike purely geometrical $f(R)$-type
reconstructions, the present approach retains the BD scalar field
as an explicit dynamical degree of freedom through the deformation
function $\alpha(z,q)$. Furthermore, recent DESI data strongly favor
modified gravity or dynamical dark energy over $\Lambda CDM$, indicating
an evolving equation of state with a phantom-to-quintessence transition.
Specific $f(R)$ models, such as exponential and logarithmic variants,
are highly consistent with these observations and statistically outperform
the standard model. These scenarios naturally mimic the required dynamical
behavior, supporting our framework's interpretation of modified gravity
as an effective description of late-time acceleration \cite{Odintsov:2025Modified,Odintsov:2025GW170817,Odintsov:2026Dynamical,Odintsov:2026ViablefR}.

Overall, the $q$-deformation modifies the effective gravitational
coupling through $\alpha\left(z,q\right)$ and leads to qualitatively
different scalar-driven expansion histories relative to standard BD
cosmology. However, the model does not by itself reproduce the observed
transition from an early decelerating era to late-time acceleration.
Additionally, a detailed assessment of observational viability, including
local-gravity constraints and potential implications for cosmological
tensions, requires extending the present vacuum analysis and is left
for future work.

\section*{Conflicts of Interest }

The authors declare no conflicts of interest.

\section*{Data Availability Statement }

No new data were created or analysed in this study.

\section*{Declaration of generative AI and AI-assisted technologies in the
writing process}

During the preparation of this work the authors used ChatGPT, Gemini
and Claude in order to improve readability of the text. After using
this tool/service, the author reviewed and edited the content as needed
and takes full responsibility for the content of the publication.

\bibliography{Q_BD_Cosm}

\end{document}